# Uniaxial plasmon polaritons *via* charge transfer at the graphene/CrSBr interface

Daniel J. Rizzo[1,*], Eric Seewald[1], Fangzhou Zhao[3], Jordan Cox[4], Kaichen Xie[5], Rocco A. Vitalone[1], Francesco L. Ruta[1,2], Daniel G. Chica[4], Yinming Shao[1], Sara Shabani[1], Evan J. Telford[1,4], Matthew C. Strasbourg[6], Thomas P. Darlington[1,6], Suheng Xu[1], Siyuan Qiu[1], Aravind Devarakonda[1,2], Takashi Taniguchi[7], Kenji Watanabe[8], Xiaoyang Zhu[4], P. James Schuck[6], Cory R. Dean[1], Xavier Roy[4,*], Andrew J. Millis[1], Ting Cao[5], Angel Rubio[3,9,10], Abhay N. Pasupathy[1,*], D.N. Basov[1,*]

[1]Department of Physics, Columbia University, New York, NY, 10027, USA

[2]Department of Applied Physics and Applied Mathematics, Columbia University, New York, NY, 10027, USA

[3]Theory Department, Max Planck Institute for Structure and Dynamics of Matter and Center for Free-Electron Laser Science, 22761 Hamburg, Germany

[4]Department of Chemistry, Columbia University, New York, NY, 10027, USA

[5]Department of Materials Science and Engineering, University of Washington, Seattle, WA 98195, USA

[6]Department of Mechanical Engineering, Columbia University, New York, NY, 10027, USA

[7]Research Center for Materials Nanoarchitectonics, National Institute for Materials Science, 1-1 Namiki, Tsukuba 305-0044, Japan

[8]Research Center for Electronic and Optical Materials, National Institute for Materials Science, 1-1 Namiki, Tsukuba 305-0044, Japan

[9]Center for Computational Quantum Physics, Flatiron Institute, New York, New York 10010, USA

[10]Nano-Bio Spectroscopy Group, Universidad del País Vasco UPV/EHU, San Sebastián 20018, Spain

*Correspondence to: db3056@columbia.edu, apn2108@columbia.edu, xr2114@columbia.edu and djr2181@columbia.edu

**Abstract**

Graphene is a privileged 2D platform for hosting confined light-matter excitations known as surface plasmon-polaritons (SPPs), as it possesses low intrinsic losses with a high degree of optical confinement. However, the inherently isotropic optical properties of graphene limit its ability to guide and focus SPPs, making it less suitable than anisotropic elliptical and hyperbolic materials as a platform for polaritonic lensing and canalization. Here, we present the graphene/CrSBr heterostructure as an engineered 2D interface that hosts highly anisotropic SPP propagation over a wide range of frequencies in the mid-infrared and terahertz. Using a combination of scanning tunneling microscopy (STM), scattering-type scanning near-field optical microscopy (s-SNOM), and first-principles calculations, we demonstrate mutual doping in excess of $10^{13}$ cm$^{-2}$ holes/electrons between the interfacial layers of graphene/CrSBr heterostructures. SPPs in graphene activated by charge transfer interact with charge-induced anisotropic intra- and interband transitions in the interfacial doped CrSBr, leading to preferential SPP propagation along the quasi-1D chains that compose each CrSBr layer. This multifaceted proximity effect both creates SPPs and endows them with anisotropic transport and propagation lengths that differ by an order-of-magnitude between the two in-plane crystallographic axes of CrSBr.

Keywords: charge transfer, graphene, CrSBr, plasmon polaritons, 2D magnetism, scanning near-field optical microscopy (SNOM), scanning tunneling microscopy (STM)

Two-dimensional (2D) van der Waals (vdW) materials are ideal atomic-scale platforms for generating confined light spanning terahertz (THz)[1-3], mid-[4-21]/near-infrared (MIR/NIR)[22] and visible[23-25] energies. These materials possess phononic,[8-11,13-16] electronic,[1-7,17,18,22] or excitonic[23-25] properties that cause the permittivity to become negative, providing the necessary conditions for hosting confined light-matter excitations known as polaritons. In general, 2D materials can support enhanced confinement, high electronic and dielectric tunability, and low losses, which has led to the realization of polaritons displaying ballistic propagation,[4] in-plane[9-11,13,23] and out-of-plane hyperbolicity,[8,15] electronically-tunable topology,[12,14] and canalization.[11,14] Among these materials, only graphene can host high-quality photonic propagation in the monolayer limit due to the uniquely high mobility of Dirac quasiparticles. Indeed, quality factors of 25 or greater have been observed for graphene surface plasmon polaritons (SPPs)[4,5] at room temperature and 150 or greater at low temperature.[4] Conversely, other vdW media typically require in excess of ~$10^2$ layers to achieve similar propagation lengths.[15] On the other hand, the isotropic nature of graphene limits its potential utility for lateral confinement and channeling of photonic modes – properties which have been observed in bulk slabs of intrinsically anisotropic materials such as α-MoO$_3$[9-13] and CrSBr.[23] The ability to impose uniaxial photonic properties on pristine sheets of graphene – free of nano-structuring and postprocessing – would enable in-plane manipulation and confinement of light propagating in an atomically-thin material.

Heterostructuring has recently been demonstrated as a viable route toward tuning the behavior of polaritons in vdW materials.[12,16-18,21] Optically-active modes in adjacent layers can be used to manipulate the dispersion of 2D polaritons and create hybrid modes.[12,19,20] In addition, emergent phenomena arising from interfacial charge transfer can significantly influence photonic behavior in 2D heterostructures. Notably, charge-transfer heterostructures (CTHs) enable non-volatile generation of SPPs in graphene,[12,17,18,21] and spatially-tunable losses of phonon-polaritons (PhPs) in hexagonal boron nitride (hBN).[16] CTHs also support nanometer-scale control of local conductivity, enabling plasmonic cavities,[21] edge plasmon polaritons,[17] and plasmonic point-scatters.[17,18] Therefore, interfacing graphene with a 2D material possessing a different work function and optical anisotropy would leverage the combined effects of charge-transfer and optical coupling to realize in-plane tunable graphene SPPs.

In this study, we demonstrate the ability to create uniaxial SPPs in graphene through heterostructuring with the air-stable vdW magnet CrSBr (shown schematically in Fig. 1A). CrSBr possesses significant in-plane structural anisotropy (Fig. 1B) that is reflected in its optical properties.[23] Each CrSBr layer is composed of 1D chains that result in a quasi-1D electronic structure.[26-29] In particular, the CrSBr $a$-axis hosts flat conduction bands (CBs) that suppress efficient electrical conductivity, while the $b$-axis CBs are more dispersive and are thus electrically conductive. Using a combination of scattering-type scanning near-field optical microscopy (s-SNOM) and scanning tunneling microscopy and spectroscopy (STM/STS), we visualize emergent electronic and nano-optical behavior arising from interfacial charge transfer and electronic anisotropy at the graphene/CrSBr interface. Our STM/STS results provide strong evidence of >0.5 eV shift in the Dirac-point energy ($E_{\text{Dirac}}$) of graphene, indicating significant charge transfer with CrSBr ($n > 10^{13}$ cm$^{-2}$) and confirming a theoretical prediction.[30] STM data also reveal topographic and density-of-states (DOS) features associated with a second-order moiré pattern, exhibiting a long-range, atomically clean interface.

We used s-SNOM to visualize the charge transfer-enabled SPPs in graphene/CrSBr. The dispersive behavior of these SPPs quantifies the magnitude of interfacial charge transfer consistent with the STS data. Our s-SNOM data further reveal a roughly order-of-magnitude difference in the SPP propagation lengths at MIR frequencies along the two in-plane crystallographic axes of CrSBr, along with a systematic suppression of the *b*-axis SPP group velocity at THz frequencies. We find that this highly anisotropic SPP behavior arises due to the quasi-1D electronic properties of the electron-doped interfacial layer of CrSBr. Here, the doped CrSBr possesses anisotropic electron-hole excitations that impart direction-dependent damping on SPPs with respect to the CrSBr in-plane crystallographic axes (Fig. 1A, B). The totality of our analysis indicates that both intra- and interband transitions play a significant role in the observed SPP anisotropy. This interpretation is supported by first-principles density functional theory (DFT) calculations, and provides a novel route for in-plane manipulation of confined light that exploits both intrinsic and emergent optical properties of vdW heterostructures.

**Atomically-Resolved Topography and Electronic Structure**

Fig. 2A shows a characteristic STM topographic image of a graphene/CrSBr heterostructure, revealing a complex pattern of short- and long-range modulations in the atomic-scale landscape. A series of Bragg peaks visible in the fast Fourier transform (FFT) of STM topography (Fig. 2A; right panel) indicates the origin of these features. These include the graphene (orange circles) and CrSBr (cyan circles) atomic lattices, and a second-order moiré pattern (yellow circles) resulting from the interference of these two layers oriented approximately 4° with respect to one another. Our ability to simultaneously image the graphene and CrSBr atomic lattices along with the second-order moiré pattern demonstrates that graphene/CrSBr forms high-quality, atomically clean interfaces with minimal structural and twist disorder.

An STS spectrum averaged over the field of view in Fig. 2A is shown in Fig. 2B. The point spectrum shows local density of states (LDOS) features derived from both graphene and the underlying CrSBr. A wide spectral tail is observe spanning negative sample biases along with broad shoulders centered roughly at 1.0 V and 2.5 V. We note that the spectra display intensity modulations likely due to moiré-induced variations in the local vacuum potential (Fig. S1). In addition, we can resolve anisotropic patterns in the vicinity of point defects, revealing electronic anisotropy that aligns with the crystallographic axes of CrSBr (Fig. S1). To isolate for LDOS features representative of graphene, we focus on sample biases spanning –0.2 to 0.8 V (Fig. 2B, inset). We observe a series of LDOS minima in the range 0 V to 0.54 V. Assigning $E_{Dirac}$ to the global minimum in the STS spectrum suggests significant interfacial charge transfer corresponding to 0.54 eV hole-doping of graphene, or $n = \sim 2 \times 10^{13}$ cm$^{-2}$ holes. Indeed, measurement of the CrSBr work function using KPFM yields a value of $W_{CrSBr} \approx 5$ eV (Fig. S2), which is larger than that of graphene ($W_{graphene} = 4.6$ eV)[31] and is consistent with interfacial hole doping of graphene and electron doping of CrSBr in graphene/CrSBr heterostructures.

**Characterization of the SPP Dispersion**

Interfacial charge transfer in graphene/CrSBr can be unambiguously quantified from s-SNOM measurements probing the frequency-dependent SPP behavior. Fig. 2C shows a map of the near-field amplitude $S_4$ at $\omega = 905$ cm$^{-1}$ collected on graphene draped over the edge of trilayer CrSBr (CrSBr edge indicated by black dashed line). While SPP fringes are typically observed near native graphene edges, the large difference in the graphene conductivity coinciding with the CrSBr edge generates a boundary capable of reflecting tip-launched SPPs and in-coupling free-space light in analogy to a true graphene edge. Thus, the observation of the characteristic fringe pattern in Fig. 2C is evidence of strong graphene-CrSBr interfacial charge transfer. We note that in Fig. 2C, the edge runs parallel to the CrSBr $b$-axis, and the observed SPPs are therefore propagating along the CrSBr $a$-axis.

To further quantify the behavior of graphene/CrSBr SPPs, we collect a series of images for frequencies spanning 860 to 1020 $cm^{-1}$. Taking the SPP line profile at each frequency (Fig. S3), we can extract the complex-valued wavevector[5]:

$$\boldsymbol{q} = q_1 + iq_2$$

where $q_1$ and $q_2$ are the real and imaginary components of $\boldsymbol{q}$, respectively. In particular, the experimental SPP dispersion $\omega(q_1)$ encodes the graphene charge carrier density and is plotted in Fig. 2D (red circles). We compare the experimental $a$-axis SPP dispersion to the calculated imaginary component of the $p$-polarized reflection coefficient, Im $r_p$, whose maxima trace the expected SPP dispersion. To calculate Im $r_p$, we input reported optical parameters,[32,33] our $a$-polarized far-field measurements of CrSBr (Fig. S4), and $E_F = 0.5$ eV for graphene (as informed by our STS measurements) (Fig. 2D). The experimental dispersion aligns well with maxima in Im $r_p$ despite there being no free parameters in our model. Thus, our near-field data validate the assignment of $E_{Dirac} = 0.54$ eV shown in Fig. 2B, confirming significant charge transfer in graphene/CrSBr. We speculate that the additional LDOS minima observed in the 0 V to 0.54 V range in Fig. 2B arise due to one or a combination of (1) Van Hove singularities in the electron doped CrSBr layer, (2) inelastic tunneling[34] and/or (3) superlattice Dirac points emerging from moiré-induced Brillouin zone folding.[35,36]

**Uniaxial SPPs**

While s-SNOM measurements of $a$-axis propagating SPPs yield several observable fringes, we find that $b$-axis SPPs have a more subtle appearance. Fig. 3A shows the corner of an exfoliated CrSBr microcrystal that is encapsulated with graphene. At this frequency (880 $cm^{-1}$), at least five observable fringes emanate along the $a$-axis from the top edge, while SPP fringes are nearly unobservable along the $b$-axis. The inset of Fig. 3A shows the average SPP line profiles along both CrSBr crystallographic axes. One faint $b$-axis fringe is observable (blue curve), while the $a$-axis curve shows multiple significant oscillations (red curve). This behavior is observed at all experimental MIR frequencies (Fig. S3), indicating SPP damping with significant in-plane anisotropy. To quantify this anisotropy, we extract the SPP $Q$-factor ($Q = \frac{q_1}{q_2}$) from the experimental SPP fringe profile (Fig. 3B). We find that SPPs have significant additional losses along the $b$-axis compared to the $a$-axis (i.e., $Q_a > Q_b$) with up to an order-of-magnitude difference in the associated $Q$-factors across the range $\omega = 850 - 1100 \text{ cm}^{-1}$. Note that $Q_a$ is comparable though smaller than nominal values obtained for SPPs measured under similar conditions in high quality $h$BN-encapsulated structures ($Q_{hBN} \approx 20$),[5] while $Q_b$ is likewise significantly suppressed. We remark that the conductivity due to Dirac electrons of graphene is isotropic. It is therefore evident that the CrSBr in our structures is acting not only as a reservoir for interfacial charge doping but also imposes anisotropic damping on graphene SPPs in the MIR that depends on the underlying orientation of the CrSBr crystallographic axes. Moreover, the observed difference in $Q_a$ and $Q_b$ is far in excess of what can be attributed to the intrinsic optical anisotropy of undoped CrSBr (Fig. S3).

**SPP Anisotropy at THz Energies**

Nano-optical measurements at THz frequencies also reveal significant anisotropic SPP behavior. Fig. 3C (left panel) shows a characteristic THz space-time map of the near-field SPP electric field ($\Delta E_{NF}$)[37] running parallel to the $a$-axis, with the CrSBr edges corresponding to the left and right edges of the panel. Extrema in the space-time map trace the SPP "worldlines" (red arrows in Fig. 3C, left panel) whose slope provide an explicit measure of the SPP group velocity, $v_g = 2\frac{\Delta x}{\Delta t}$, as recently discovered.[37] Extracting the $a$-axis group velocity $v_g^a$ from our space-time maps yields an average value of $22 \pm 3$ μm/ps. Space-time maps collected along the $b$-axis also show SPP worldlines (Fig. 3C, right panel) whose associated group velocity $v_g^b$ is $18 \pm 3$ μm/ps. Histograms of the measured $a$- and $b$-axis group velocities can be found in Fig. 3D, revealing a systematic reduction of $v_g^b$ compared to $v_g^a$. Thus, our space-time maps demonstrate that proximity-induced suppression of $b$-axis SPP propagation extends from the MIR down to THz energies.

**Mechanism for Uniaxial SPPs**

In order to understand the microscopic origins of SPP anisotropy, we perform first-principles DFT calculations on model heterostructures consisting of graphene on monolayer CrSBr (Fig. 4B) and graphene on bilayer CrSBr (Fig. S5A–C). The monolayer-on-monolayer band structure is plotted in Fig. 4B (see Fig. S5 for monolayer-on-bilayer and spin-polarized band structures). Characteristic features of the isolated graphene and CrSBr layers can be clearly identified, such as the graphene Dirac cone and the quasi-1D structure of the CrSBr CB. Notably, the theoretical value of $E_{\text{Dirac}}$ is approximately 0.5 eV indicating significant hole-doping of the graphene layer – in good agreement with both tunneling and nano-optical experiments (Fig. 2B, D), as well as previous theoretical predictions.[30] We note that most of the charge (>80% according to our calculations) transferred from graphene is found to be localized at the top-most CrSBr layer due to the formation of a potential gradient within CrSBr. Furthermore, the inferred $E_{\text{Dirac}} = 0.5$ eV in the graphene/bilayer CrSBr calculations (Fig. S5D) is mirrored in the graphene/monolayer CrSBr analysis (Fig. 4B), further indicating that the charge transfer is confined to the interfacial CrSBr layer. Indeed, other charge-transfer heterostructures are predicted to possess charge localization primarily in the interfacial layers due to similar effects.[38]

Our DFT calculations also show that the conduction band of CrSBr is now significantly electron doped and can thus support new intra- and interband transitions. Notably, free-carriers associated with the conduction band of doped CrSBr are highly anisotropic ($m_a = m_{\Gamma\rightarrow X} = 3.1m_e$ and $m_b = m_{\Gamma\rightarrow Y} = 0.2m_e$) indicating that doped CrSBr is much more conductive along the $b$-axis than along the $a$-axis. Hence, the associated anisotropic Drude response of electron-doped CrSBr is likely to impart direction-dependence of both plasmonic group velocity and damping. In the case of interband transitions, the conduction band of doped CrSBr resides in close proximity to another unoccupied band (Fig. 4A) whose energy separation is anisotropic and overlaps with the SPP energies probed in our experiment. Thus, emergent interband transitions accessed through electron doping CrSBr will likely also induce anisotropy in the SPP response of graphene/CrSBr.

To explore the respective roles of intra- and interband transitions on the creation of uniaxial SPPs, we next investigate the origin of direction-dependent surface plasmon damping from first-principles calculations. We evaluate the overall plasmon scattering rate $\gamma$:

$$\gamma(E, \boldsymbol{q}) \sim \int \delta(E - [E_c(\boldsymbol{k} + \boldsymbol{q}) - E_v(\boldsymbol{k})]|\langle\psi_{c,\boldsymbol{k}+\boldsymbol{q}}|H_{int}|\psi_{v,\boldsymbol{k}}\rangle|^2)d^3\boldsymbol{k}$$

(1)

Where $E_c$ and $E_v$ are the energies of unoccupied and occupied states, respectively, $\boldsymbol{k}$ are the crystal momenta, $\boldsymbol{q}$ the SPP momenta, and $H_{int}$ represents the electron-photon interaction Hamiltonian. The form of $\gamma$ contains the joint density of states (JDOS) and matrix elements of $H_{int}$. Therefore, we can use the JDOS to quantify the anisotropic plasmon scattering rate by sampling electronic transitions whose crystal momenta match the momenta of the SPPs in the MIR ($\sim 2 - 8 \times 10^5$ cm$^{-1}$). Here we consider two model structures: (1) free-standing monolayer CrSBr that has been electron-doped to match the charge transfer with graphene (Fig. 4A), and (2) a graphene/monolayer CrSBr heterostructure (Fig. 4B). In so doing we aim to parse the separate effects of charge doping and the presence of the graphene layer on the calculated scattering rate.

Along the *a*-axis, the calculated JDOS for the free-standing doped CrSBr shows two peaks at ~0.0 eV and ~0.14 eV (red curves, Fig. 4C) that mirror JDOS peaks in graphene/monolayer CrSBr at ~0.03 eV and ~0.20 eV (red curves, Fig. 4D). For both structures, the lower energy peaks mainly come from intraband transitions in the two closely packed flat bands near the Fermi level along $\Gamma - X$, while the higher energy peaks mainly come from interband transitions between these bands. We note that for monolayer CrSBr/graphene, there are additional closely packed bands due to zone folding (Fig. 4B). However, we exclude these transitions in our calculations since $\langle \psi_{c,\boldsymbol{k}+\boldsymbol{q}} | H_{int} | \psi_{v,\boldsymbol{k}} \rangle$ is negligible when the momentum conservation is only made possible due to zone folding. Thus, the contribution to the scattering rate from the graphene layer is modest. Notably, both model structures show a JDOS minimum (i.e., suppressed plasmon damping) along the *a*-axis around 0.1 eV, corresponding to the energy range covered in our MIR experiments.

In contrast, the JDOS along the *b*-axis (blue curves, Fig. 4C,D) reveals a series of peaks that are roughly 10 times higher than the *a*-axis JDOS around 0.1 eV. Here, the *b*-axis peak is composed of both intraband and interband transitions in roughly equal proportion. Since the quality factor scales with the ratio between the plasma frequency and the scattering rate ($Q \sim \frac{\omega_p}{\gamma}$)[17] and the *a*- and *b*-axis plasma frequencies are similar (Fig. S3), our calculated damping ratio explains the measured order-of-magnitude reduction of the $Q$ factor along the *b*-axis compared to the *a*-axis for SPPs in the MIR. The notion of an enhanced *b*-axis scattering rate is also consistent with our nano-THz data. We remark that the group velocity of low-momentum THz plasmons is in fact impacted by the scattering rate as described in ref. [37]. The reduction of the *b*-axis THz group velocity by 30% in Fig. 3C,D is thus in accord with the scenario of increased scattering rate. The totality of our analyses of SPPs in the MIR and THz regimes reveal that anisotropic SPP propagation in graphene/CrSBr is a cooperative effect of strong electronic anisotropy in CrSBr and proximal charge transfer with graphene.

**Outlook**

We have performed a multi-modal STM and s-SNOM experimental and theoretical study of graphene/CrSBr heterostructures, revealing significant proximity-induced reciprocal charge transfer and electronically-mediated plasmonic anisotropy. By leveraging sensitivity to both the local electronic and nano-optical behavior of graphene/CrSBr, we unravel the subtle interplay of intrinsic optical properties, electronic anisotropy, and emergent plasmonic damping. Combined with theoretical insights provided by first-principles calculations, our observations yield a complete mechanistic picture for engineering uniaxial SPPs in graphene, and enable 2D manipulation of polaritons in an atomically-thin material.

Our results have significant implications for the manipulation of graphene SPPs and 2D polaritons in general, and outline a novel approach for inducing anisotropy in intrinsically isotropic media. Indeed, previously proposed routes for achieving uniaxial or anisotropic SPPs rely on complex device fabrication[12] or intrinsic anisotropy in the SPP host-medium[39] – significantly limiting the potential material platforms for engineering 2D light. Using our proximity-based approach, it is now possible to leverage uniaxial graphene SPPs to direct plasmonically-mediated energy-transfer in 2D with nanoscale precision – enabling 2D lensing, waveguiding, and canalization in the monolayer limit. Moreover, the behavior observed in this study is generic to graphene-based charge transfer heterostructures composed of anisotropic semiconducting building blocks (e.g., black phosphorous, $ReSe_2$). We also anticipate that this behavior can be further manipulated through *in situ* modulation of the graphene/CrSBr chemical potential and twist-engineering multiple graphene-CrSBr interfaces. In addition, it is likely that other interlayer effects can be leveraged in graphene/CrSBr to tune directional SPP transport, including anisotropic shake-off bands,[40] moiré-induced Brillouin zone folding[35,36], and quasi-1D charge density wave formation. Finally, we foresee opportunities for electronic and plasmonic manipulation of 2D magnetically-ordered phases in CrSBr using our charge-transfer platform and vice versa.

## Methods

### *Device Fabrication*

Single crystals of CrSBr were synthesized using a chemical vapor transport reaction with source and sink zone temperatures of 930°C and 850°C, respectively. The details of the synthesis can be found in ref. [41]. We note that CrSBr naturally cleaves along the principal *a*- and *b*-axes, with a high *a*- to *b*-axis aspect ratio in resulting microcrystals. As such, the relative orientation of graphene plasmon propagation with respect to the underlying CrSBr can be readily identified from the macroscopic features of the stack; this assignment was confirmed with polarized photoluminescence experiments (Fig. S6).

For s-SNOM devices, flakes of few-layer CrSBr, graphene, and few-layer *h*BN (< 8 nm) were obtained via mechanical exfoliations. Flake thicknesses were identified optically and confirmed with atomic force microscopy (AFM). The *h*BN/graphene/CrSBr heterostructure was prepared with the dry stamp method.[42] A transparent polydimethylsiloxane (PDMS) cube (~1 × 1 × 1 mm$^3$) was covered by a thin polycarbonate (PC) polymer film and used to pick up the top *h*BN layer. After picking up the graphene layers with the *h*BN, the heterostructure was transferred onto the trilayer CrSBr at elevated temperatures (~120– 180°C). The sample surface was then washed with chloroform, acetone, and isopropyl alcohol to remove the melted PC polymer. To further remove residual transfer polymer and interfacial bubbles, the surfaces of $h$BN /graphene/CrSBr heterostructures were imaged with approximately 1 nN of contact force using contact-mode AFM with line spacing of <100 nm.

For STM experiments, exfoliated flakes were instead picked up in the following sequence: $h$BN, CrSBr, then graphene. Polymer-supported stacks were then flipped and deposited onto a Si/SiO$_2$ substrate with no further processing. Electrical contact was established to the graphene layer by microsoldering using Field's metals.[43]

### Scanning Tunneling Microscopy and Spectroscopy

STM/STS measurements were performed in a home-built, ultra-high vacuum system at 5.7K. Atomically sharp tips were electrochemically etched from tungsten wire and spectroscopically calibrated using Shockley surface states on single crystal Au(111).[44] Multiple independently prepared tips were used to verify the accuracy and reproducibility of the measurements.

### Scanning Near-field Optical Microscopy

The MIR s-SNOM and KPFM measurements were performed on a commercial Neaspec system under ambient conditions using commercial Arrow$^{TM}$ AFM probes with nominal resonant frequencies of $f$ = 75 kHz or 256 kHz. Tunable continuous wave quantum cascade lasers produced by Daylight Solutions were used spanning wavelengths from 9 to 11.7 μm. The detected signal was demodulated at the fourth harmonic of the tip tapping frequency in order to minimize far-field contributions to the scattered light. The fourth harmonic of the near-field scattering amplitude (S$_4$) and phase ($\Phi_4$) were collected simultaneously using a pseudoheterodyne interferometry technique.

The THz s-SNOM measurements we performed on a commercial Neaspec system under ambient conditions using AFM tips produced by Rocky Mountain Nanotechnology, LLC with a nominal resonant frequency of 30 – 80 kHz. The THz broadband pulse is generated and detected using a pair of photo-conductive antennas (PCAs, Menlo Systems GmbH). The THz radiation from the PCA emitter is collimated by a TPX lens and focused onto the tip and sample by a parabolic mirror. The scattered field is detected by an unbiased PCA in the time domain using a 50 fs, 780 nm gate beam, and the photocurrent signal is demodulated by a lock-in amplifier at harmonics of the tip tapping frequency.

### *Ab-initio Calculations of Graphene/CrSBr Heterostructures*

First principles calculations were performed utilizing DFT implemented in the Quantum ESPRESSO package.[45] Norm-conserving pseudopotentials were employed alongside a plane-wave energy cutoff of 90 Ry.[46] For structural relaxation, the spin-polarized Perdew-Burke-Ernzerhof exchange-correlation functional was employed with van der Waals corrections (PBE-D3).[47] The structures were fully relaxed until the force on each atom was <0.005 eV/Å. In monolayer CrSBr, the lattice constants along the $a$ and $b$ axes were determined to be 3.54 and 4.72 Å, respectively. In monolayer graphene, the lattice constant $a$ was relaxed to 2.465 Å. The graphene/CrSBr heterostructure was constructed with a supercell of 5 × 2 graphene rectangular conventional cell containing 4 carbon atoms stacked atop a 6 × 1 CrSBr supercell, aligning the $a$ and $b$ axes of CrSBr bilayer along the armchair and zigzag directions of graphene monolayer, respectively. The graphene monolayer experiences compressive strain of ~0.5% along the armchair direction, and compressive strain of ~4% along the zigzag direction. The vdW spatial gap between the top CrSBr layer (from top Br atoms) and the graphene monolayer is 3.46 Å. A vacuum region of 15 Å was added in the out-of-plane direction to avoid interaction between periodic images. Brillouin zone sampling in the graphene/CrSBr heterostructure was performed using an 8 × 30 × 1 $k$-grid. Dipole correction was applied in all calculations of the graphene/CrSBr heterostructure.[48] A Gaussian smearing of 1 meV was adopted for electron occupation. The first-principles calculation of JDOS for the estimation of plasmon damping is performed on a $24 \times 96 \times 1$ $k$-grid for the graphene/monolayer CrSBr heterostructure with supercell, and on a $144 \times 96 \times 1$ $k$-grid for the free-standing monolayer CrSBr calculation with doping, so the smallest sampled wavevector for plasmon damping is $2 \times 10^5$ $cm^{-1}$ along both $a$ and $b$ axes. The Brillouin zone unfolding in the JDOS calculation of CrSBr/graphene supercell utilizes the BandUp code.[49,50] The RPA dielectric function calculations are performed using the BerkeleyGW code.[51] The inverse of the dielectric function is calculated on a $12 \times 48 \times 1$ $k$-grid, with a cutoff energy of 8.0 Ry, and a slab truncation of Coulomb interaction. The smallest sampled plasmon wavevector in the dielectric function calculation is $4 \times 10^5$ $cm^{-1}$ along both $a$ and $b$ axes, which is limited by the large computational cost due to large supercells. The frequency dependence of the inverse dielectric function is calculated by the Adler-Wiser[52,53] formalism implemented in the BerkeleyGW code, with an energy broadening of 5 meV. The inverse dielectric function shows the plasmon peaks with linewidths on the order of 20 meV, with larger linewidth for the plasmon mode along the $b$-axis.

## Acknowledgements

Research on charge transfer plasmons is solely supported as part of the Energy Frontier Research Center on Programmable Quantum Materials funded by the U.S. Department of Energy (DOE), Office of Science, Basic Energy Sciences (BES), under Award No DE-SC0019443. K.W. and T.T. acknowledge support from the JSPS KAKENHI (Grant Numbers 20H00354, 21H05233 and 23H02052) and World Premier International Research Center Initiative (WPI), MEXT, Japan. The first-principles calculations (Fig. S5) are based upon work supported by the National Science Foundation under Award No. DMR-2339995. F. Z. acknowledges the support of the Alexander von Humboldt-Stiftung for the financial support from the Humboldt Research Fellowship. Synthesis of the CrSBr crystals was in part supported by the National Science Foundation (NSF) through the Columbia Materials Science and Engineering Research Center on Precision-Assembled Quantum Materials (DMR-2011738). T.C. utilized advanced computational, storage, and networking infrastructure provided by the Hyak supercomputer system and funded by the University of Washington Molecular Engineering Materials Center at the University of Washington (NSF MRSEC DMR-2308979). A.D. acknowledges support from the Simons Foundation Society of Fellows program (Grant No. 855186).

**Author Contributions**

D.J.R. performed all s-SNOM measurements and analysis. E.S. and S.S. performed the STM experiments and analysis. F.Z. and A.R. performed DFT calculations of the plasmonic properties and analysis of plasmon lifetime. K.X. and T.C. performed DFT calculations and analysis. J.C., E.J.T., and A.D. fabricated heterostructures for s-SNOM and STM experiments. R.A.V. and S.X. aided THz imaging and analysis. F.L.R. modeled the near-field data. Y.S. and S.Q. performed KPFM and far-field optical characterization of CrSBr. D.G.C. performed growth and characterization of CrSBr single crystals. M.C.S. and T.P.D. performed PL measurements. K.W. and T.T. performed growth and characterization of $h$BN single crystals. A.R., P.J.S., and X.Z. advised and participated in experimental design and interpretation. C.R.D. and X.R. advised synthesis and device fabrication efforts. A.N.P. advised STM experiments. D.N.B. advised s-SNOM experiments. The manuscript was written with input from all authors.

**Competing Interests**

The authors declare no competing financial interests.

**Data Availability**

All data presented in the manuscript are available upon request.

**Supporting Information Available**

Supporting Information contains PL data, ancillary STS and s-SNOM data, CrSBr FTIR spectra, KPFM measurements, and additional details for optical modeling and DFT calculations.

## References

1    Bandurin, D. A. *et al.* Resonant terahertz detection using graphene plasmons. *Nature Communications* **9**, 5392 (2018). https://doi.org/10.1038/s41467-018-07848-w

2    Alonso-González, P. *et al.* Acoustic terahertz graphene plasmons revealed by photocurrent nanoscopy. *Nature nanotechnology* **12**, 31-35 (2017). https://doi.org/10.1038/nnano.2016.185

3    Low, T. & Avouris, P. Graphene Plasmonics for Terahertz to Mid-Infrared Applications. *ACS Nano* **8**, 1086-1101 (2014). https://doi.org/10.1021/nn406627u

4    Ni, G. X. *et al.* Fundamental limits to graphene plasmonics. *Nature* **557**, 530-533 (2018). https://doi.org/10.1038/s41586-018-0136-9

5    Woessner, A. *et al.* Highly confined low-loss plasmons in graphene–boron nitride heterostructures. *Nature Materials* **14**, 421-425 (2015). https://doi.org/10.1038/nmat4169

6    Chen, J. *et al.* Optical nano-imaging of gate-tunable graphene plasmons. *Nature* **487**, 77-81 (2012).

7    Fei, Z. *et al.* Gate-tuning of graphene plasmons revealed by infrared nano-imaging. *Nature* **487**, 82-85 (2012).

8    Dai, S. *et al.* Tunable Phonon Polaritons in Atomically Thin van der Waals Crystals of Boron Nitride. *Science* **343**, 1125-1129 (2014). https://doi.org/10.1126/science.1246833

9    Ma, W. *et al.* In-plane anisotropic and ultra-low-loss polaritons in a natural van der Waals crystal. *Nature* **562**, 557-562 (2018). https://doi.org/10.1038/s41586-018-0618-9

10   Zheng, Z. *et al.* A mid-infrared biaxial hyperbolic van der Waals crystal. *Science Advances* **5**, eaav8690  https://doi.org/10.1126/sciadv.aav8690

11   Duan, J. *et al.* Multiple and spectrally robust photonic magic angles in reconfigurable α-MoO3 trilayers. *Nature Materials* **22**, 867-872 (2023). https://doi.org/10.1038/s41563-023-01582-5

12   Ruta, F. L. *et al.* Surface plasmons induce topological transition in graphene/α-MoO3 heterostructures. *Nature Communications* **13**, 3719 (2022). https://doi.org/10.1038/s41467-022-31477-z

13   Zheng, Z. *et al.* Highly Confined and Tunable Hyperbolic Phonon Polaritons in Van Der Waals Semiconducting Transition Metal Oxides. *Advanced Materials* **30**, 1705318 (2018). https://doi.org:https://doi.org/10.1002/adma.201705318

14   Li, P. *et al.* Collective near-field coupling and nonlocal phenomena in infrared-phononic metasurfaces for nano-light canalization. *Nature Communications* **11**, 3663 (2020). https://doi.org/10.1038/s41467-020-17425-9

15   Ni, G. *et al.* Long-Lived Phonon Polaritons in Hyperbolic Materials. *Nano Letters* **21**, 5767-5773 (2021). https://doi.org/10.1021/acs.nanolett.1c01562

16   Rizzo, D. J. *et al.* Polaritonic Probe of an Emergent 2D Dipole Interface. *Nano Letters* **23**, 8426-8435 (2023). https://doi.org/10.1021/acs.nanolett.3c01611

17   Rizzo, D. J. *et al.* Charge-Transfer Plasmon Polaritons at Graphene/α-RuCl3 Interfaces. *Nano Letters* **20**, 8438-8445 (2020). https://doi.org/10.1021/acs.nanolett.0c03466

18   Rizzo, D. J. *et al.* Nanometer-Scale Lateral p–n Junctions in Graphene/α-RuCl3 Heterostructures. *Nano Letters* **22**, 1946-1953 (2022). https://doi.org/10.1021/acs.nanolett.1c04579

19   Dai, S. *et al.* Graphene on hexagonal boron nitride as a tunable hyperbolic metamaterial. *Nature Nanotechnology* **10**, 682-686 (2015). https://doi.org/10.1038/nnano.2015.131

20      Brar, V. W. *et al.* Hybrid Surface-Phonon-Plasmon Polariton Modes in Graphene/Monolayer h-BN Heterostructures. *Nano Letters* **14**, 3876-3880 (2014). https://doi.org/10.1021/nl501096s

21      Kim, B. S. Y. *et al.* Ambipolar charge-transfer graphene plasmonic cavities. *Nature Materials* **22**, 838-843 (2023). https://doi.org/10.1038/s41563-023-01520-5

22      Shao, Y. *et al.* Infrared plasmons propagate through a hyperbolic nodal metal. *Science Advances* **8**, eadd6169  https://doi.org/10.1126/sciadv.add6169

23      Ruta, F. L. *et al.* Hyperbolic exciton polaritons in a van der Waals magnet. *Nature Communications* **14**, 8261 (2023). https://doi.org/10.1038/s41467-023-44100-6

24      Sternbach, A. J. *et al.* Femtosecond exciton dynamics in WSe2 optical waveguides. *Nature Communications* **11**, 3567 (2020). https://doi.org/10.1038/s41467-020-17335-w

25      Hu, F. *et al.* Imaging exciton–polariton transport in MoSe2 waveguides. *Nature Photonics* **11**, 356-360 (2017). https://doi.org/10.1038/nphoton.2017.65

26      Klein, J. *et al.* The Bulk van der Waals Layered Magnet CrSBr is a Quasi-1D Material. *ACS Nano* **17**, 5316-5328 (2023). https://doi.org/10.1021/acsnano.2c07316

27      Wu, F. *et al.* Quasi-1D Electronic Transport in a 2D Magnetic Semiconductor. *Advanced Materials* **34**, 2109759 (2022). https://doi.org:https://doi.org/10.1002/adma.202109759

28      Wilson, N. P. *et al.* Interlayer Electronic Coupling on Demand in a 2D Magnetic Semiconductor. *arXiv preprint arXiv:2103.13280* (2021).

29      Ziebel, M. E. *et al.* CrSBr: An Air-Stable, Two-Dimensional Magnetic Semiconductor. *Nano Letters* **24**, 4319-4329 (2024). https://doi.org/10.1021/acs.nanolett.4c00624

30      Xie, K., Zhang, X.-W., Xiao, D. & Cao, T. Engineering Magnetic Phases of Layered Antiferromagnets by Interfacial Charge Transfer. *ACS Nano* **17**, 22684-22690 (2023). https://doi.org/10.1021/acsnano.3c07125

31      Yu, Y.-J. *et al.* Tuning the Graphene Work Function by Electric Field Effect. *Nano Letters* **9**, 3430-3434 (2009). https://doi.org/10.1021/nl901572a

32      Caldwell, J. D. *et al.* Sub-diffractional volume-confined polaritons in the natural hyperbolic material hexagonal boron nitride. *Nature Communications* **5**, 5221 (2014). https://doi.org/10.1038/ncomms6221

33      Kučírková, A. & Navrátil, K. Interpretation of Infrared Transmittance Spectra of SiO2 Thin Films. *Appl. Spectrosc.* **48**, 113-120 (1994).

34      Zhang, Y. *et al.* Giant phonon-induced conductance in scanning tunnelling spectroscopy of gate-tunable graphene. *Nature Physics* **4**, 627-630 (2008). https://doi.org/10.1038/nphys1022

35      Yankowitz, M. *et al.* Emergence of superlattice Dirac points in graphene on hexagonal boron nitride. *Nature Physics* **8**, 382-386 (2012). https://doi.org/10.1038/nphys2272

36      Li, Y. *et al.* Anisotropic band flattening in graphene with one-dimensional superlattices. *Nature Nanotechnology* **16**, 525-530 (2021). https://doi.org/10.1038/s41565-021-00849-9

37      Xu, S. *et al.* Electronic interactions in Dirac fluids visualized by nano-terahertz spacetime mapping. *arXiv preprint arXiv:2311.11502* (2023).

38      Biswas, S., Li, Y., Winter, S. M., Knolle, J. & Valentí, R. Electronic Properties of $\ensuremath{\alpha}\text{\ensuremath{-}}{\mathrm{RuCl}}_{3}$ in Proximity to Graphene. *Physical Review Letters* **123**, 237201 (2019). https://doi.org/10.1103/PhysRevLett.123.237201

39    Low, T. *et al.* Plasmons and Screening in Monolayer and Multilayer Black Phosphorus. *Physical Review Letters* **113**, 106802 (2014). https://doi.org/10.1103/PhysRevLett.113.106802

40    Ulstrup, S. *et al.* Observation of interlayer plasmon polaron in graphene/WS2 heterostructures. *Nature Communications* **15**, 3845 (2024). https://doi.org/10.1038/s41467-024-48186-4

41    Bae, Y. J. *et al.* Exciton-coupled coherent magnons in a 2D semiconductor. *Nature* **609**, 282-286 (2022). https://doi.org/10.1038/s41586-022-05024-1

42    Wang, L. *et al.* One-Dimensional Electrical Contact to a Two-Dimensional Material. *Science* **342**, 614-617 (2013). https://doi.org/10.1126/science.1244358

43    Girit, Ç. Ö. & Zettl, A. Soldering to a single atomic layer. *Applied Physics Letters* **91**, 193512 (2007). https://doi.org/10.1063/1.2812571

44    Chen, W., Madhavan, V., Jamneala, T. & Crommie, M. F. Scanning Tunneling Microscopy Observation of an Electronic Superlattice at the Surface of Clean Gold. *Physical Review Letters* **80**, 1469-1472 (1998). https://doi.org/10.1103/PhysRevLett.80.1469

45    Giannozzi, P. *et al.* QUANTUM ESPRESSO: a modular and open-source software project for quantum simulations of materials. *Journal of Physics: Condensed Matter* **21**, 395502 (2009). https://doi.org/10.1088/0953-8984/21/39/395502

46    Hamann, D. R. Optimized norm-conserving Vanderbilt pseudopotentials. *Physical Review B* **88**, 085117 (2013). https://doi.org/10.1103/PhysRevB.88.085117

47    Grimme, S. Semiempirical GGA-type density functional constructed with a long-range dispersion correction. *Journal of Computational Chemistry* **27**, 1787-1799 (2006). https://doi.org:https://doi.org/10.1002/jcc.20495

48    Bengtsson, L. Dipole correction for surface supercell calculations. *Physical Review B* **59**, 12301-12304 (1999). https://doi.org/10.1103/PhysRevB.59.12301

49    Medeiros, P. V. C., Stafström, S. & Björk, J. Effects of extrinsic and intrinsic perturbations on the electronic structure of graphene: Retaining an effective primitive cell band structure by band unfolding. *Physical Review B* **89**, 041407 (2014). https://doi.org/10.1103/PhysRevB.89.041407

50    Medeiros, P. V. C., Tsirkin, S. S., Stafström, S. & Björk, J. Unfolding spinor wave functions and expectation values of general operators: Introducing the unfolding-density operator. *Physical Review B* **91**, 041116 (2015). https://doi.org/10.1103/PhysRevB.91.041116

51    Deslippe, J. *et al.* BerkeleyGW: A massively parallel computer package for the calculation of the quasiparticle and optical properties of materials and nanostructures. *Computer Physics Communications* **183**, 1269-1289 (2012). https://doi.org:https://doi.org/10.1016/j.cpc.2011.12.006

52    Wiser, N. Dielectric Constant with Local Field Effects Included. *Physical Review* **129**, 62-69 (1963). https://doi.org/10.1103/PhysRev.129.62

53    Adler, S. L. Quantum Theory of the Dielectric Constant in Real Solids. *Physical Review* **126**, 413-420 (1962). https://doi.org/10.1103/PhysRev.126.413

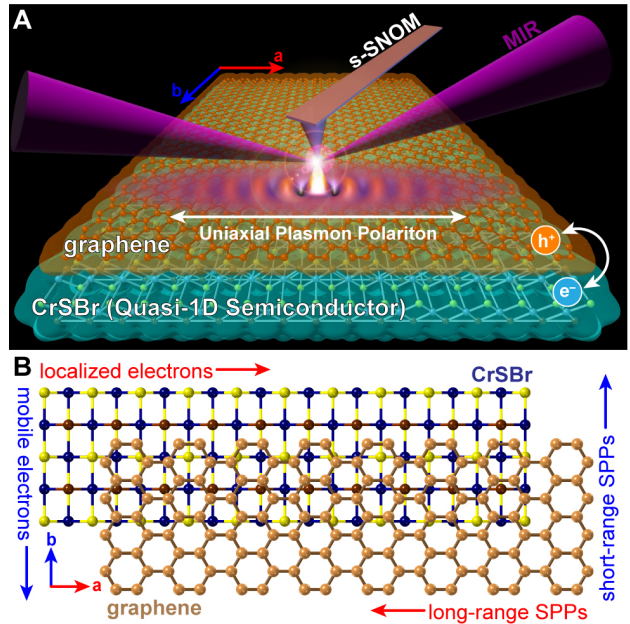

**Figure 1. Schematic of charge transfer and uniaxial plasmon-polariton propagation in graphene/CrSBr heterostructures.** (**A**) Overview of s-SNOM measurements in graphene/CrSBr heterostructures. The difference in work functions between these layers leads to hole-doped graphene and electron-doped CrSBr. Uniaxial surface plasmon-polaritons (SPPs) are generated upon illumination of the AFM tip with mid-infrared (MIR) light. (**B**) CrSBr displays a highly anisotropic crystal structure, forming 1D chains along the in-plane *a*-axis. The resulting electronic structure is quasi-1D, with *b*-axis carriers possessing a much higher mobility than *a*-axis carriers. Proximate interactions between graphene SPPs and the underlying CrSBr carriers leads to preferential SPP propagation along the 1D *a*-axis chains.

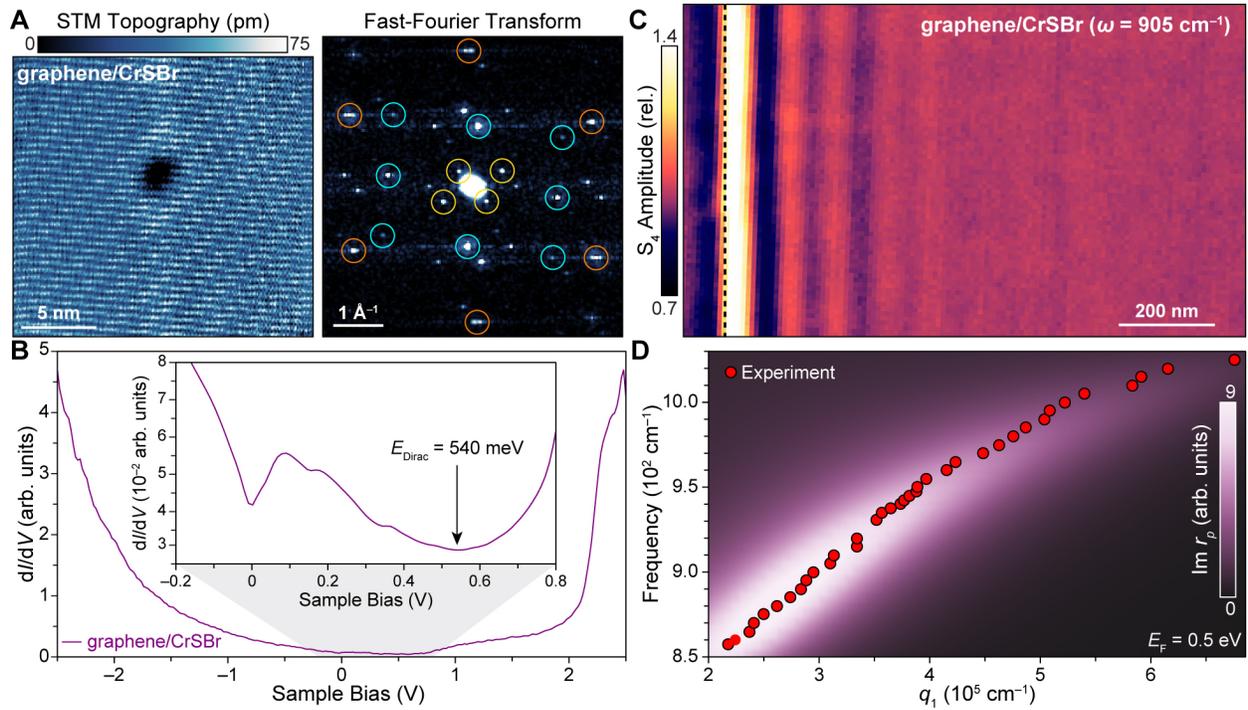

**Figure 2. Multi-modal characterization of interfacial charge transfer in graphene/CrSBr heterostructures.** (**A**) Left panel: Atomically-resolved topographic STM image of a graphene/CrSBr heterostructure ($V_S$ = 1.2 V, $I$ = 50 pA, $T$ = 5.7 K). Right panel: Fast Fourier transform (FFT) of topographic data shows Bragg peaks associated with the graphene atomic lattice (orange circles), the CrSBr atomic lattice (cyan circles) and the second-order moiré pattern (yellow circles) corresponding to a twist angle of ~4° between graphene and CrSBr. (**B**) STS collected on graphene/CrSBr heterostructure ($V_S$ = 1.8 V, $I$ = 50 pA). Inset: Low bias STS ($V_S$ = 0.3 V, $I$ = 100 pA) shows a d$I$/d$V$ minimum at 540 mV corresponding to the Dirac-point energy ($E_{Dirac}$) of graphene shifted due to interfacial charge transfer with the underlying CrSBr. Additional nearby d$I$/d$V$ minima are also observed at 0 V, 150 mV, and 330 mV. (**C**) Typical s-SNOM image of a graphene/CrSBr heterostructure showing oscillations in the near-field $S_4$ amplitude that are characteristic of SPPs ($\omega$ = 905 cm$^{-1}$; $S_4$ normalized relative to the value in the graphene/CrSBr bulk). The graphene is draped over the CrSBr edge (dashed black line), creating a sharp gradient in the graphene charge density that acts as a hard boundary for plasmonic reflections. (**D**) Red circles: The experimental SPP dispersion for graphene/CrSBr extracted from the line profiles of SPP fringes collected at different frequencies (see Fig. S3). Calculated Im $r_p$ for the experimental stack using an input value of $E_F$ = 0.5 eV for the graphene Fermi energy. Maxima in the loss function correspond well with the experimental dispersion, indicating that the SPP behavior is consistent with an 0.5 eV shift in $E_{Dirac}$ of graphene due to charge transfer with the underlying CrSBr.

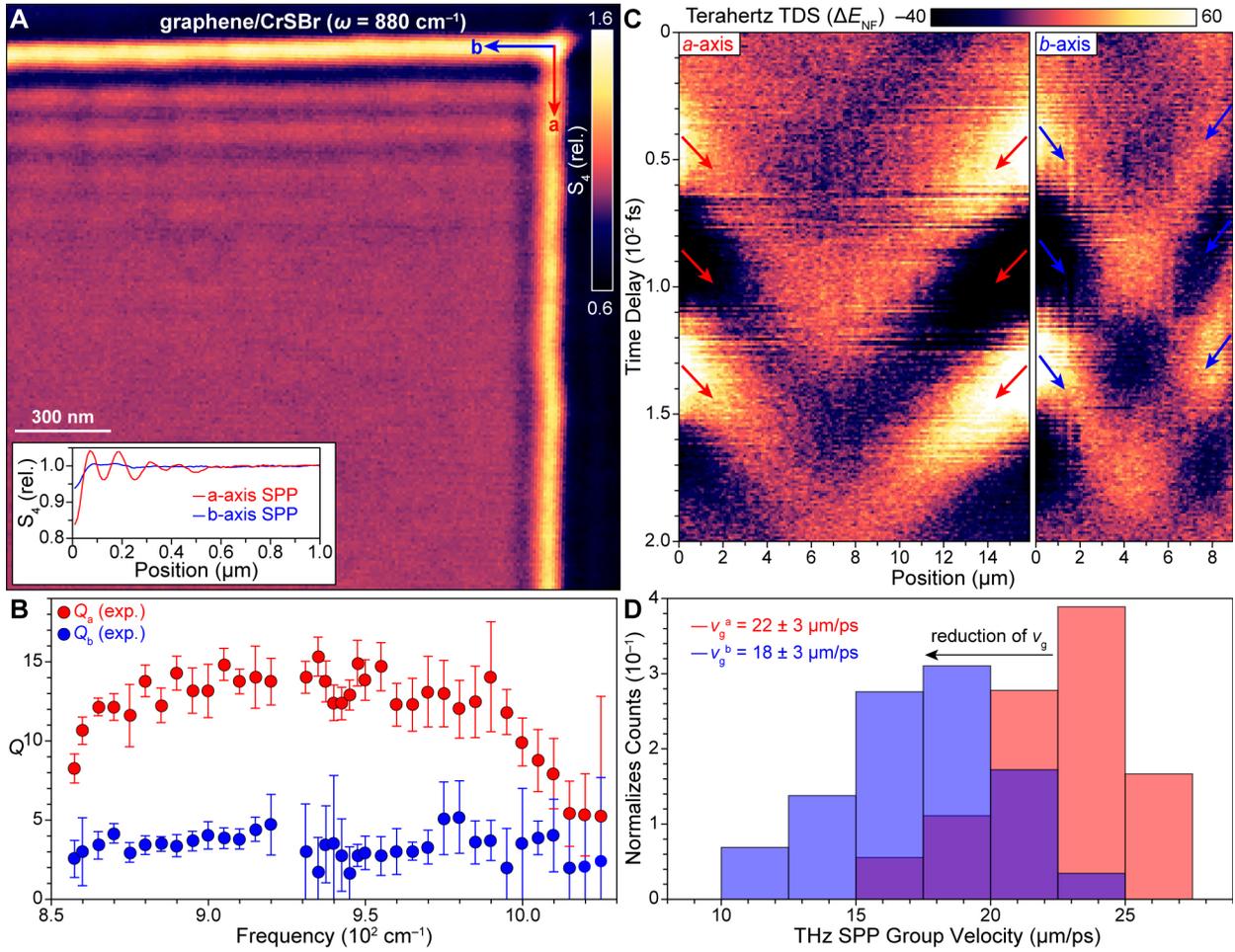

**Figure 3. Proximity-induced anisotropy and uniaxial SPPs in graphene/CrSBr heterostructures.** (**A**) Map of the near-field $S_4$ amplitude in a graphene/CrSBr heterostructure showing multiple SPP fringes propagating along the CrSBr $a$-axis, while $b$-axis fringes are highly suppressed. Inset: The average line profile of SPP fringes along the $a$-axis (red curve) versus the $b$-axis (blue curve) shows a significantly diminished decay length for the latter ($\omega = 880$ cm$^{-1}$; $S_4$ normalized relative to the value in the graphene/CrSBr bulk). (**B**) The experimentally-extracted frequency-dependent $Q$-factor for SPPs propagating along the $a$-axis (red circles) and $b$-axis (blue circles). (**C**) Left panel: Space-time map of the near-field plasmon electric field ($\Delta E_{NF}$) conducted along the $a$-axis of a graphene/CrSBr heterostructure (See Fig. S7 for background subtraction procedure). The edges of the images correspond to the edges of CrSBr. The red arrows indicate extrema that correspond to the worldlines of propagating SPPs. The group velocity of $a$-axis modes $v_g^a$ is twice the slope of the worldline. Right panel: The same as the left panel but for $b$-axis propagating modes with blue arrows indicating the SPP worldlines with group velocity $v_g^b$. (**D**) Histograms of the associated values of $v_g^a$ and $v_g^b$ extracted from $N \geq 18$ worldlines for each direction showing a systematic suppression of $v_g^b$ compared to $v_g^a$.

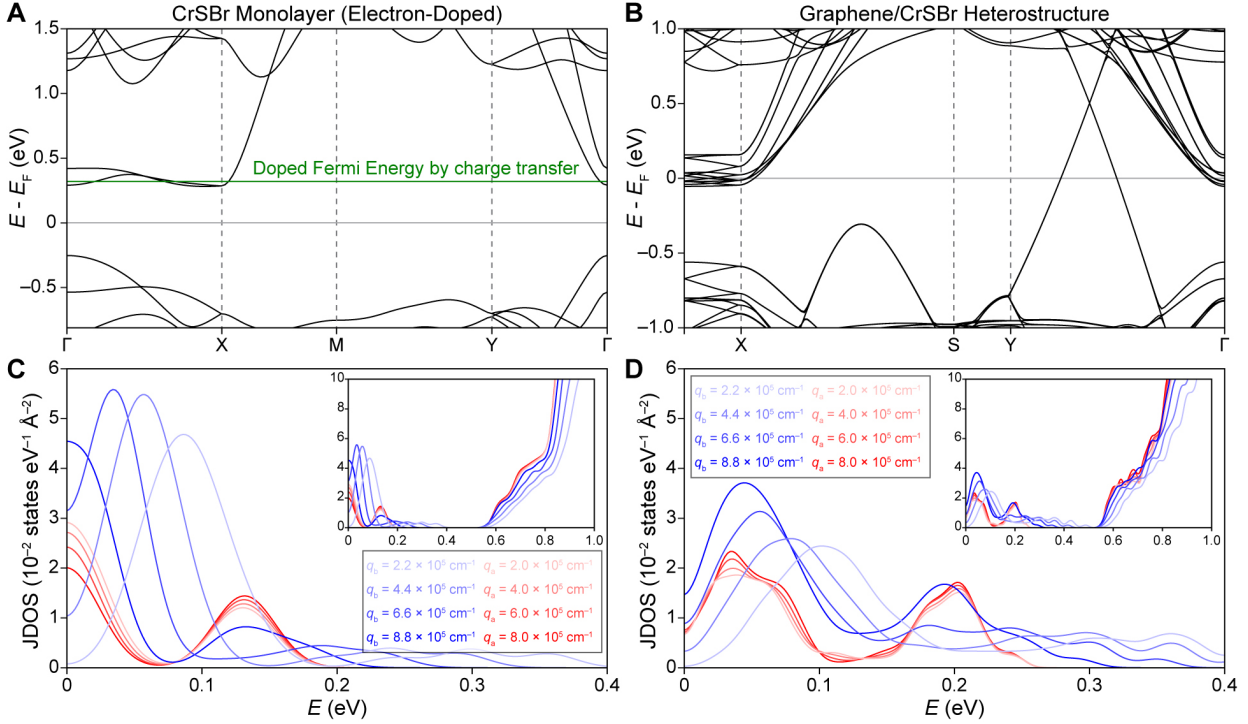

**Figure 4. First-principles calculation of the JDOS for electron-doped CrSBr monolayer and graphene/CrSBr monolayer heterostructures.** (**A**) Band structure of free-standing monolayer CrSBr calculated by DFT with PBE functional. The solid green line indicates the Fermi level of the CrSBr due to charge transfer with graphene (The charge transfer is ~0.03 electrons per CrSBr unit cell). (**B**) Band structure of graphene/CrSBr monolayer heterostructure calculated by DFT with PBE functional. (**C**) The calculated joint density of state (JDOS) for an electron-doped CrSBr monolayer with the Fermi level marked by the solid green line in panel (A). The JDOS corresponds to the plasmon damping transition with wavevector $\boldsymbol{q}$ and transition energy $E$ from the occupied and unoccupied manifolds, defined by $JDOS(E, \boldsymbol{q}) = \int \delta(E - [E_c(\boldsymbol{k} + \boldsymbol{q}) - E_v(\boldsymbol{k})]) d^3\boldsymbol{k}$. Red curves denote the JDOS corresponding to the plasmon damping transition in the $a$ direction ($\boldsymbol{q} = \boldsymbol{q}_a$), and blue curves the $b$ direction ($\boldsymbol{q} = \boldsymbol{q}_b$). Curves with different hues denote different $\boldsymbol{q}$ values in the two directions as indicated in the plot legend. (**D**) The calculated JDOS for the CrSBr monolayer/graphene heterostructure. The band structure of the heterostructure exhibits significant band folding, though the transitions between band folded states are unlikely to influence the plasmon damping process. Therefore, the plotted quantity is $\gamma(E, \boldsymbol{q}) \sim \int \delta(E - [E_c(\boldsymbol{k} + \boldsymbol{q}) - E_v(\boldsymbol{k})]) |\langle \psi_{c,\boldsymbol{k}+\boldsymbol{q}}|H_{int}|\psi_{v,\boldsymbol{k}}\rangle|^2) d^3\boldsymbol{k}$, where $H_{int}$ is the electron-photon interaction in the plasmon damping, and $\langle \psi_{c,\boldsymbol{k}+\boldsymbol{q}_a}|H'|\psi_{v,\boldsymbol{k}}\rangle$ is negligible for the zone folded bands. The insets in both (C) and (D) show the JDOS over a larger energy range. The anisotropy in JDOS is significant below ~0.4 eV, where the transitions are mostly from the flat bands near the Fermi level in the $a$ direction, and from the very dispersive bands near the Fermi level in the $b$ direction, respectively. At higher energy the JDOS is quite isotropic since the transitions involve the more isotropic band complex above 1.0 eV and below -0.5 eV in panel (A).

**Supporting Information**

**Uniaxial plasmon polaritons *via* charge transfer at the graphene/CrSBr interface**

**Table of Contents:**

**Calculating $Q$ for the experimental stack**

To calculate the expected $Q$-factors explicitly, we must account for all layers in our vdW stack using the following approximation:[1]

$$Q^{-1} = \frac{q_2}{q_1} \approx \frac{\sigma_1}{\sigma_2} + \frac{\epsilon_2^{eff}}{\epsilon_1^{eff}} = \frac{\gamma_g(\omega)}{\omega} + \frac{\epsilon_2^{eff}}{\epsilon_1^{eff}}$$

(S1)

where $\sigma = \sigma_1 + i\sigma_2$ is the complex optical conductivity of graphene, $\gamma_g(\omega)$ is the frequency-dependent graphene scattering rate,[1] and $\epsilon^{eff} = \epsilon_1^{eff} + i\epsilon_2^{eff}$ is the effective permittivity of the material surrounding graphene. Here, $\epsilon^{eff}$ depends on the dielectric functions of the media surrounding graphene (but not graphene itself), and is composed of both substrate and superstrate components. Diagram S1 shows a schematic of our experimental stack and Tables S1–S4 show the associated oscillator parameters for each layer.

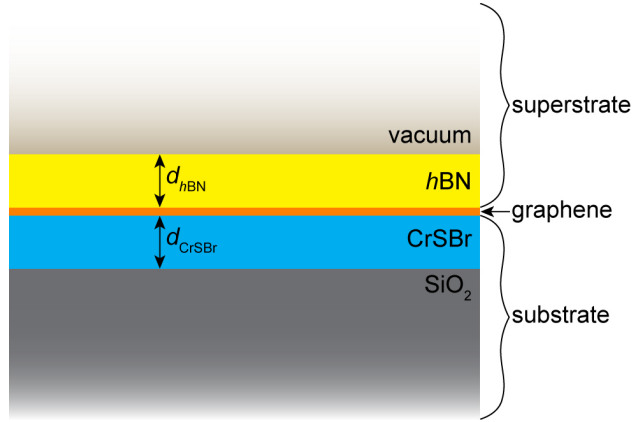

**Diagram S1. Schematic of experimental heterostructure used for optical modeling.**

Here, the in-plane ($a = xy$) and out-of-plane ($a = z$) responses of $h$BN were modeled with single-oscillator Drude-Lorentz forms:

$$\epsilon_{hBN}^a(\omega) = \epsilon_\infty^a + \epsilon_\infty^a \frac{(\omega_{LO}^a)^2 - (\omega_{TO}^a)^2}{(\omega_{TO}^a)^2 - \omega^2 - i\gamma^a\omega}$$

(S2)

And the SiO₂ and CrSBr layers were modeled as the sum of multiple Lorentzians:

$$\epsilon(\omega) = \epsilon_\infty + \sum_i \frac{\Omega_i}{\omega_i^2 - \omega^2 - i\gamma_i\omega}$$

(S3)

Where the *a*- and *b*-axis permittivites of CrSBr have difference oscillator parameters.

The superstrate is composed of a semi-infinite layer of vacuum and an 8.7-nm-thick layer of *h*BN, and the substrate consists of a semi-infinite layer of SiO$_2$ with a 2.4-nm-thick layer of CrSBr. The effective superstrate permittivity, $\epsilon_{sup}^{eff}$ is given by[2]

$$\epsilon_{sup}^{eff} = \epsilon_{hBN}^z \frac{\epsilon_0 + \epsilon_{hBN}^z \tanh(-ik_{sup}d_{hBN})}{\epsilon_{hBN}^z + \epsilon_0 \tanh(-ik_{sup}d_{hBN})}, \quad k_{sup} = i\frac{\sqrt{\epsilon_{hBN}^{xy}}}{\sqrt{\epsilon_{hBN}^z}}q .$$

(S4)

Where $d_{hBN}$ is the thickness of the *h*BN layer, $\epsilon_{hBN}^{xy}$ and $\epsilon_{hBN}^z$ are the in-plane and out-of-plane components of the *h*BN permittivity, and $\epsilon_0 = 1$ is the vacuum permittivity. The effective superstrate permittivity, $\epsilon_{sub}^{eff}$ can be similarly addressed by substituting $\epsilon_{hBN} \rightarrow \epsilon_{CrSBr}$, $\epsilon_0 \rightarrow \epsilon_{SiO2}$, and $d_{hBN} \rightarrow d_{CrSBr}$:

$$\epsilon_{sub}^{eff} = \epsilon_{CrSBr}^z \frac{\epsilon_{SiO2} + \epsilon_{CrSBr}^z \tanh(-ik_{sub}^j d_{CrSBr})}{\epsilon_{CrSBr}^z + \epsilon_{SiO2} \tanh(-ik_{sub}^j d_{CrSBr})}, \quad k_{sub}^j = i\frac{\sqrt{\epsilon_{CrSBr}^j}}{\sqrt{\epsilon_{CrSBr}^z}}q .$$

(S5)

Where *j* = *a* or *b* depending on whether we are solving for the effective permittivity along the *a* or *b* axis of CrSBr. The overall effective permittivity $\epsilon^{eff}$ is the average of the superstrate and substrate permittivities:

$$\epsilon^{eff}(\omega, q) = (\epsilon_{\text{sub}} + \epsilon_{\text{sup}})/2$$

(S6)

For Eq. (S1), we evaluate $\epsilon^{eff}(\omega, q)$ at momenta $q_1$ given by the energy-moment dispersion extracted from maxima in Im $r_p$ such that $\epsilon^{eff} \equiv \epsilon^{eff}(\omega, q_1)$.

| | $\omega_{\text{TO}}$ (cm$^{-1}$) | $\omega_{\text{LO}}$ (cm$^{-1}$) | $\gamma$ (cm$^{-1}$) | $\epsilon_\infty$ |
|---|---|---|---|---|
| $\epsilon^{xy}$ | 1360 | 1614 | 7 | 4.9 |
| $\epsilon^z$ | 760 | 825 | 2 | 2.95 |

**Table S1: Oscillator parameters for $h$BN[3] .**

| $i$ | $\omega_i$ (cm$^{-1}$) | $\Omega_i$ (cm$^{-1}$) | $\gamma_i$ (cm$^{-1}$) |
|---|---|---|---|
| 1 | 1270 | 224.76$i$ | 216 |
| 2 | 1205 | 208.47 | 78 |
| 3 | 1072 | 865.50 | 49 |
| 4 | 802.3 | 310.28 | 80 |

**Table S2: Oscillator parameters for SiO$_2$. The high-frequency permittivity is $\epsilon_\infty = 1.96$. SiO$_2$ parameters derived from ref. [4]**

| $i$ | $\omega_i$ (cm$^{-1}$) | $\Omega_i$ (cm$^{-1}$) | $\gamma_i$ (cm$^{-1}$) |
|---|---|---|---|
| 1 | 501.41 | 2143.41 | 0 |
| 2 | 771.73 | 239.80 | 43.54 |
| 3 | 838.63 | 928.04 | 298.45 |
| 4 | 1240.73 | 2851.03 | 1195.43 |
| 5 | 2028.01 | 3413.46 | 1343.38 |
| 6 | 2881.99 | 3456.49 | 1557.61 |

**Table S3: Oscillator parameters for CrSBr $a$-axis. The high-frequency permittivity for the CrSBr $a$-axis is $\epsilon_\infty^a = 14.45$.**

| $i$ | $\omega_i$ (cm$^{-1}$) | $\Omega_i$ (cm$^{-1}$) | $\gamma_i$ (cm$^{-1}$) |
|---|---|---|---|
| 1 | 110.31 | 3325.26 | 0 |
| 2 | 676.12 | 113.59 | 18.86 |
| 3 | 705.53 | 325.61 | 52.77 |
| 4 | 786.22 | 831.29 | 201.00 |
| 5 | 1019.84 | 2043.03 | 691.51 |
| 6 | 2925.39 | 3823.45 | 2692.67 |
| 7 | 1566.03 | 3081.89 | 1550.68 |

**Table S4: Oscillator parameters for CrSBr $b$-axis. The high-frequency permittivity for the CrSBr $b$-axis is $\epsilon_\infty^b = 17.72$.**

## Ab-initio Calculations of Graphene/Bilayer CrSBr Heterostructure

First principles calculations were performed utilizing DFT implemented in the Quantum ESPRESSO package.[5] Norm-conserving pseudopotentials were employed alongside a plane-wave energy cutoff of 85 Ry.[6] For structural relaxation, the spin-polarized Perdew-Burke-Ernzerhof exchange-correlation functional was employed with van der Waals corrections (PBE-D2).[7] The structures were fully relaxed until the force on each atom was <0.005 eV/Å. In bilayer CrSBr, the lattice constants along the $a$ and $b$ axes were determined to be 3.51 and 4.71 Å, respectively, and the interlayer distance (Cr-Cr) was calculated to be 8.09 Å. In monolayer graphene, the lattice constant $a$ was relaxed to 2.457 Å. The graphene/bilayer CrSBr heterostructure was constructed with an 8 × 2 graphene supercell stacked atop a 5 × 1 CrSBr supercell, aligning the $a$ and $b$ axes of CrSBr bilayer along the armchair and zigzag directions of graphene monolayer, respectively. The graphene monolayer experiences tensile strain of ~3% along the armchair direction, and compressive strain of ~4% along the zigzag direction. The vdW spatial gap between the top CrSBr layer (from top Br atoms) and the graphene monolayer is 3.38 Å. A vacuum region of 15 Å was added in the out-of-plane direction to avoid interaction between periodic images. Brillouin zone sampling in the graphene/CrSBr heterostructure was performed using an 8 × 30 × 1 $k$-grid. Dipole correction was applied in all calculations for graphene/CrSBr heterostructures.[8] A Gaussian smearing of 1 meV was adopted for electron occupation. We note that the ground state of bilayer CrSBr changes from AFM to FM in our calculations, with a small energy difference < 0.1 meV/(unit cell-layer) between the phases. Calculations using the AFM ground states do not substantially change the amount of charge transferred or the CrSBr effective mass compared with the FM calculations (Fig. S6).

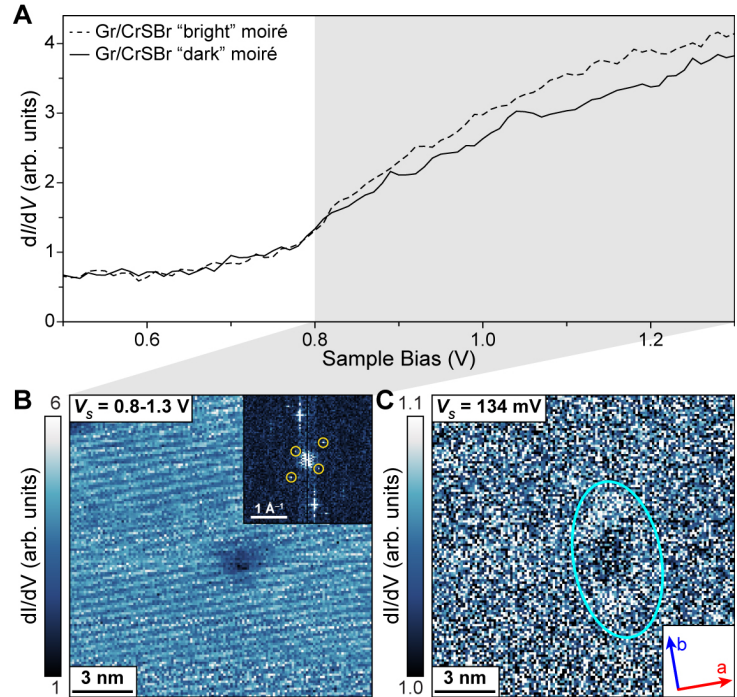

**Figure S1. Defect scattering and moiré-dependence of STS on graphene/CrSBr.** (**A**) $dI/dV$ point spectra collected at different locations within the second-order moiré formed at the graphene/CrSBr interface. Spectra collecting in the "bright" moiré regions (dashed black line) display systematically greater spectral intensity compared to those in "dark" moiré regions (solid black line) over sample biases 0.8– 1.3 V. (**B**) $dI/dV$ map showing integrated spectral intensity over sample biases 0.8 V through 1.3 V. A periodic modulation in the spectral intensity commensurate with the second-order moiré pattern is evident. Inset: FFT of the integrated $dI/dV$ map showing Bragg peaks associated with the second-order moiré pattern (yellow circles). (**C**) $dI/dV$ map taken at $V_s$ = 134 mV showing LDOS associated with anisotropic quasiparticle scattering that registers with the underlying $a$- and $b$-axes of CrSBr.

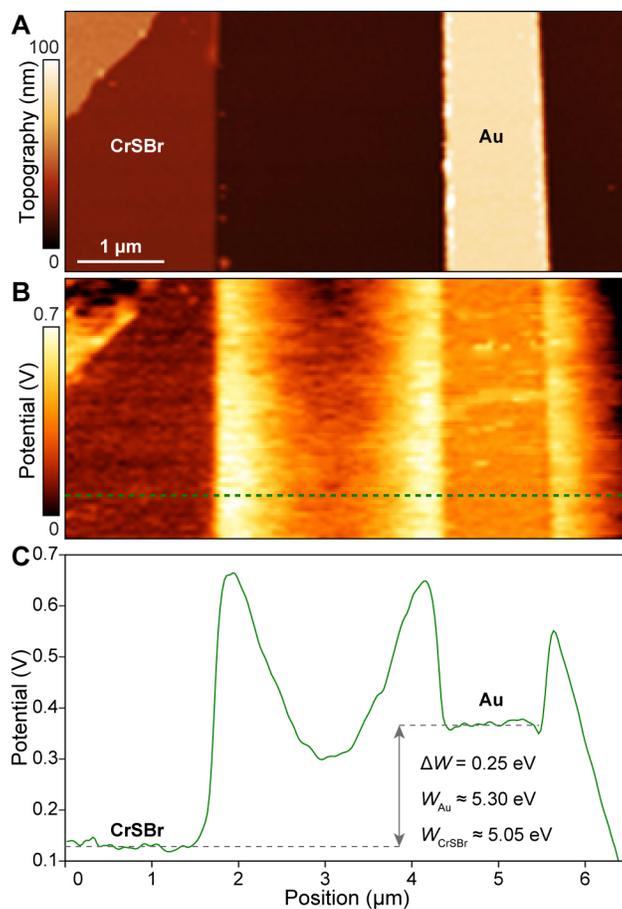

**Figure S2. Kelvin probe force microscopy of CrSBr.** (**A**) AFM topography of a CrSBr microcrystal next to a Au contact. (**B**) KPFM image of the region shown in panel (A). (**C**) Average linecut of the contact potential measured along the dashed green line in panel (B). The difference in the work functions of Au and CrSBr is measured to be 0.25 eV. Given a nominal value of 5.30 eV for a vacuum-deposited gold thin film,[9] the work function of CrSBr is measured to be 5.05 eV.

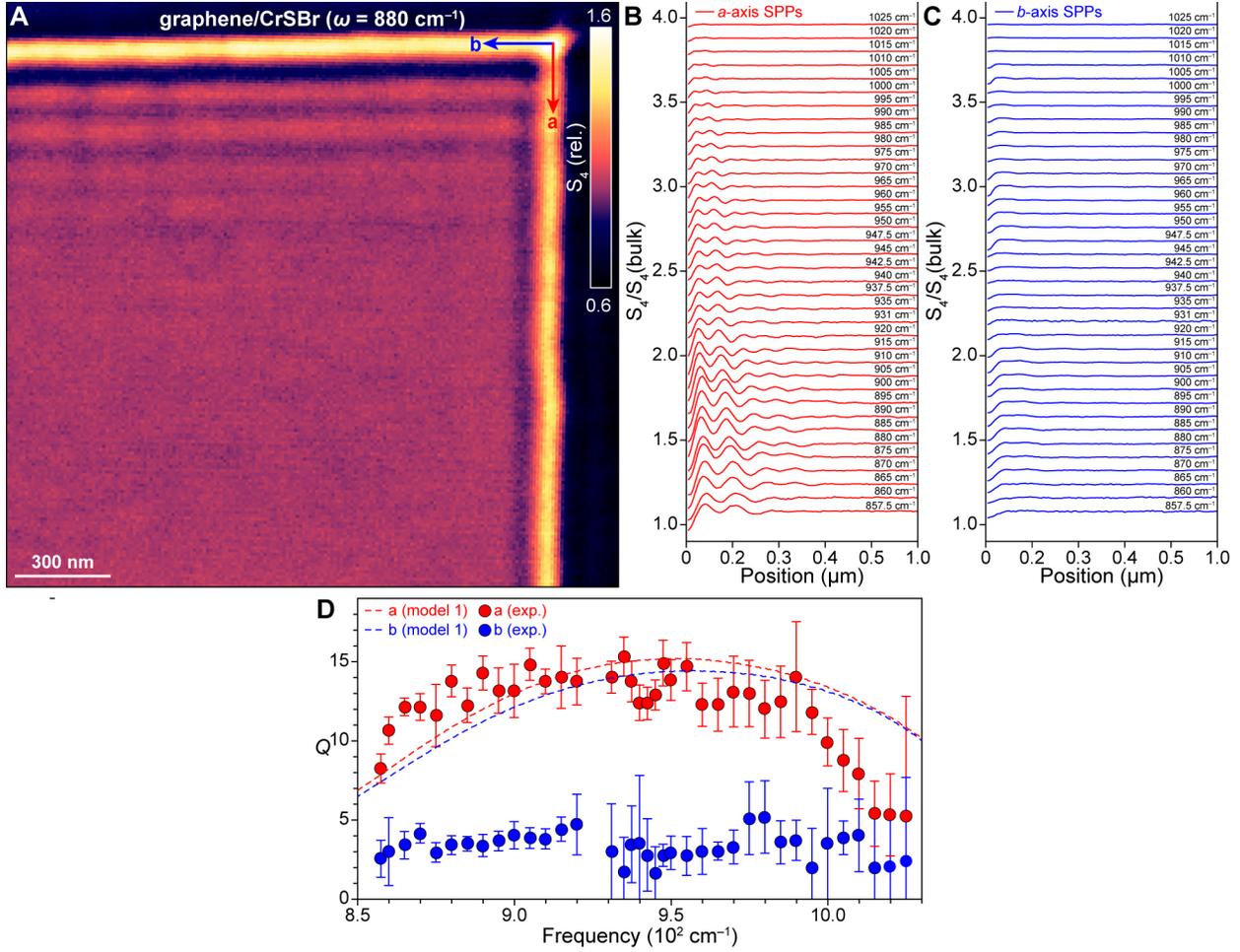

**Figure S3. Frequency-dependent near-field measurements of graphene/CrSBr.** (**A**) Map of the near-field $S_4$ amplitude in a graphene/CrSBr heterostructure reproduced from Fig. 2 of the main manuscript ($\omega$ = 880 cm$^{-1}$; $S_4$ normalized relative to the value in the graphene/CrSBr bulk). (**B**) Average line profile for plasmonic fringes running parallel to the CrSBr *a*-axis for frequencies ranging from $\omega$ = 857.5 cm$^{-1}$ to 1025 cm$^{-1}$. (**C**) Same as (B) but for fringes running parallel to the CrSBr *b*-axis. The values of *Q* were extracted by fitting the line profiles in (B) and (C) to the functional ansatz $S_0 + A\frac{e^{-iqx}}{R^a+x^a} + BH_0^{(1)}(2qx)$,[2,10] where $S_0$ is the bulk near-field amplitude, *R* is the approximate tip radius (25 nm), $H_0^{(1)}$ is the first Hankel function of order zero, *A* and *B* are sample angle- and tip-dependent scaling factors, respectively, and *a* is a geometric factor ≈0.1. For *b*-axis line profiles with $\omega$ > 920 cm$^{-1}$, *q* was constrained to be equal to that extracted from the associated *a*-axis profile. (**D**) The experimentally-extracted frequency-dependent *Q*-factor for SPPs propagating along the *a*-axis (red circles) and *b*-axis (blue circles). The expected values of $Q_a$ (dashed red curve) and $Q_b$ (dashed blue curve) are plotted based on intrinsic optical parameters.

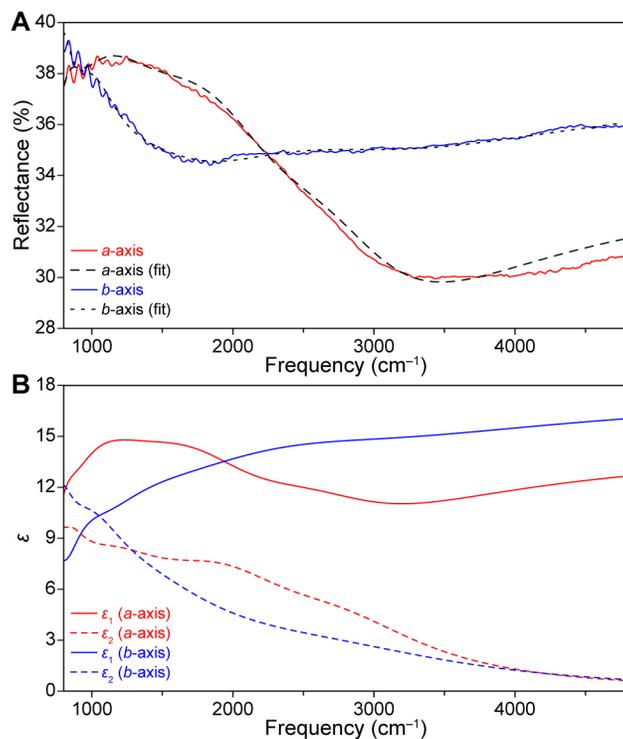

**Figure S4. CrSBr FTIR spectra and associated permittivity.** (**A**) Far-field FTIR spectra collected on a SiO₂-supported CrSBr microcrystal for light polarized along the *a*-axis (red line) and *b*-axis (blue line). The dashed and dotted black lines show the computed reflectance based on the best fit optical parameters. (**B**) The *a*- (red lines) and *b*- (blue lines) components of the real (solid lines) and imaginary (dashed lines) permittivity extracted from the FTIR spectra in panel (A).

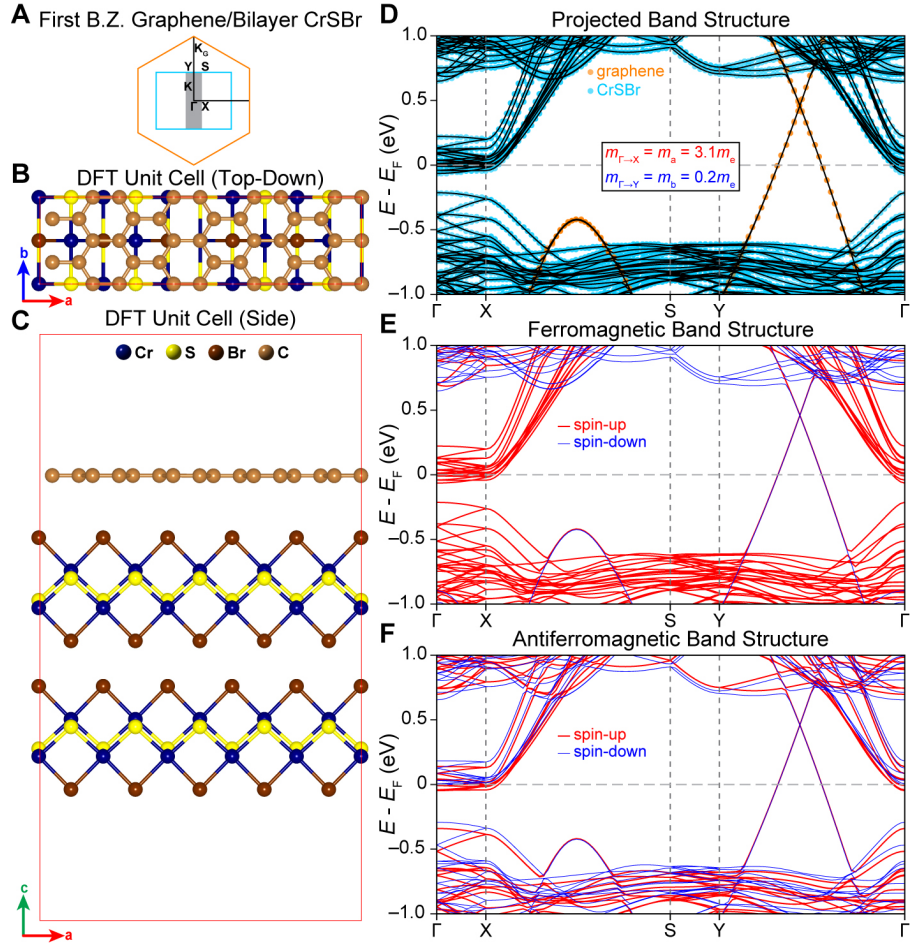

**Figure S5. Full DFT supercell and spin-polarized band structures of graphene/bilayer CrSBr.** (**A**) Schematic of the first Brillouin zone of graphene (orange hexagon), CrSBr (cyan rectangle), and the supercell shown in (B) and (C) (solid gray rectangle). High symmetry points associated with the isolated layers and the composite structure are labelled. (**B**) Top-down and (**C**) side views of the DFT supercell used for calculations in panels (D)– (F). The red line indicates the boundaries of the supercell. (**D**) Band structure for the graphene/bilayer CrSBr supercell shown in (B) and (C). States localized in the graphene (CrSBr) layer are indicated with orange (cyan) circles. A shift of ~0.5 eV in $E_{\mathrm{Dirac}}$ is observed for the hole-doped graphene layer. The CrSBr conduction band is observed to be electron-doped, with *a*-axis carriers possessing an order-of-magnitude greater effective mass ($m_{\Gamma \to X} = m_a = 3.1 m_e$) than *b*-axis carriers ($m_{\Gamma \to Y} = m_b = 0.2 m_e$). (**E**) Spin-up (red lines) and spin-down (blue) band structure for the supercell shown in panels (B) and (C) given interlayer ferromagnetic ordering in the CrSBr bilayer. (**F**) Same as (E) but for interlayer antiferromagnetic ordering in the CrSBr bilayer.

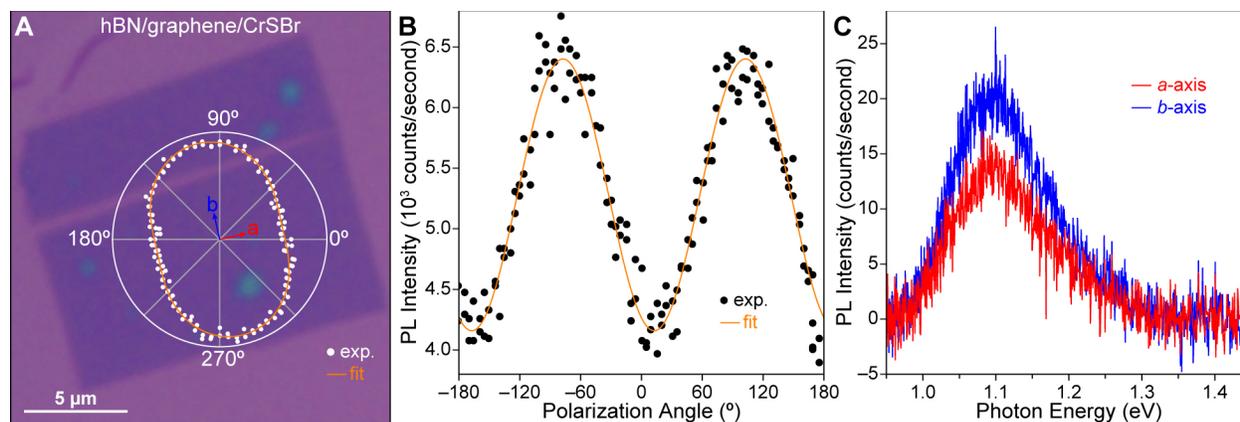

**Figure S6. Polarized photoluminescence of *h*BN/graphene/CrSBr.** (**A**) Optical overview of van der Waals heterostructure consisting of *h*BN/graphene/CrSBr. The CrSBr microcrystal naturally exfoliates such that the *a*-to-*b*-axis aspect ratio is greater than one. A polar plot of the PL intensity is overlaid on the CrSBr microcrystal, indicating that the PL intensity is at a maximum along the *b*-axis and minimum along the *a*-axis. (**B**) Plot of the angle-dependent PL intensity reproduced from panel (**A**). (**C**) Characteristic PL spectra taken along the CrSBr *a*- and *b*-axes. Compared to PL in isolated few-layer CrSBr,[11] the PL spectrum is red-shifted by ~200 meV and presents with less angular anisotropy.

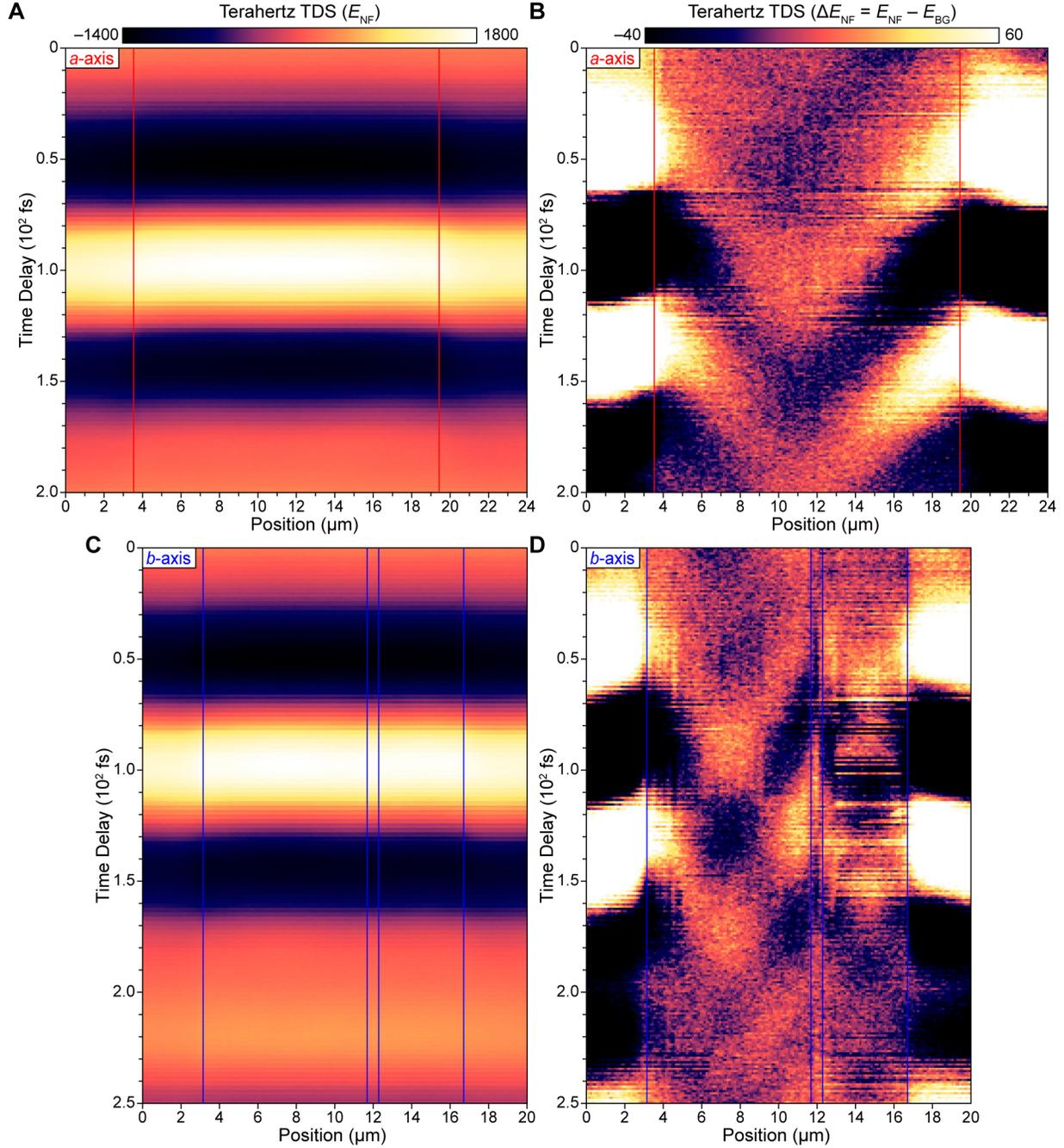

**Figure S7. Background subtraction of THz space-time maps.** (**A**) Raw space-time map of the near-field electric field $E_{NF}$ collected along the *a*-axis of a graphene/CrSBr heterostructure. The signal is dominated by the waveform of the THz pulse. (**B**) The space-time map shown in (A) with the background THz pulse removed, leaving just the near-field plasmon electric field ($\Delta E_{NF} = E_{NF} - E_{BG}$). Here, each line is leveled by subtracting the mean signal within the CrSBr flake (CrSBr edges indicated by the red vertical lines). (**C**) Same as (A) but conducted along the CrSBr *b*-axis. (**D**) Same as (B) but collected along the CrSBr *b*-axis. We note the presence of a crack in the middle of the flake, giving rise to plasmonic reflection at four boundaries instead two.

| **(A)** $q_a$ or $q_b$ (unit: $\times 10^5 cm^{-1}$) | **2** | **4** | **6** | **8** |
|---|---|---|---|---|
| $\Gamma(E = 0.1\ eV, q = q_a)$ (unit: states/eV/Å$^2$) | 0.00323 | 0.00314 | 0.00310 | 0.00306 |
| $\Gamma(E = 0.1\ eV, q = q_b)$ (unit: states/eV/Å$^2$) | 0.00258 | 0.00591 | 0.02860 | 0.04527 |
| **(B)** $q_a$ or $q_b$ (**unit:** $\times 10^5 cm^{-1}$) | **2** | **4** | **6** | 8 |
| $\Gamma(E = 0.1\ eV, q = q_a)$ (unit: states/eV/Å$^2$) | 0.00153 | 0.00220 | 0.00288 | 0.00326 |
| $\Gamma(E = 0.1\ eV, q = q_b)$ (unit: states/eV/Å$^2$) | 0.00969 | 0.01326 | 0.01971 | 0.02505 |

**Table S5. The calculated joint density of state (JDOS) for plasmon damping. (A)** The calculated JDOS for plasmon damping for free-standing monolayer CrSBr with the Fermi level shifted due to the charge transfer between the interfacial CrSBr layer and graphene. **(B)** The calculated JDOS for plasmon damping for the CrSBr monolayer/graphene heterostructure. All the calculated quantities are integrated over a 0.02 eV energy window centered at 0.1 eV.

# References

1     Ni, G. X. *et al.* Fundamental limits to graphene plasmonics. *Nature* **557**, 530-533 (2018). https://doi.org:10.1038/s41586-018-0136-9

2     Rizzo, D. J. *et al.* Charge-Transfer Plasmon Polaritons at Graphene/α-RuCl3 Interfaces. *Nano Letters* **20**, 8438-8445 (2020). https://doi.org:10.1021/acs.nanolett.0c03466

3     Caldwell, J. D. *et al.* Sub-diffractional volume-confined polaritons in the natural hyperbolic material hexagonal boron nitride. *Nature Communications* **5**, 5221 (2014). https://doi.org:10.1038/ncomms6221

4     Fei, Z. *et al.* Infrared Nanoscopy of Dirac Plasmons at the Graphene–SiO2 Interface. *Nano Letters* **11**, 4701-4705 (2011). https://doi.org:10.1021/nl202362d

5     Giannozzi, P. *et al.* QUANTUM ESPRESSO: a modular and open-source software project for quantum simulations of materials. *Journal of Physics: Condensed Matter* **21**, 395502 (2009). https://doi.org:10.1088/0953-8984/21/39/395502

6     Hamann, D. R. Optimized norm-conserving Vanderbilt pseudopotentials. *Physical Review B* **88**, 085117 (2013). https://doi.org:10.1103/PhysRevB.88.085117

7     Grimme, S. Semiempirical GGA-type density functional constructed with a long-range dispersion correction. *Journal of Computational Chemistry* **27**, 1787-1799 (2006). https://doi.org:https://doi.org/10.1002/jcc.20495

8     Bengtsson, L. Dipole correction for surface supercell calculations. *Physical Review B* **59**, 12301-12304 (1999). https://doi.org:10.1103/PhysRevB.59.12301

9     Sachtler, W. M. H., Dorgelo, G. J. H. & Holscher, A. A. The work function of gold. *Surface Science* **5**, 221-229 (1966). https://doi.org:https://doi.org/10.1016/0039-6028(66)90083-5

10     Woessner, A. *et al.* Highly confined low-loss plasmons in graphene–boron nitride heterostructures. *Nature Materials* **14**, 421-425 (2015). https://doi.org:10.1038/nmat4169

11     Wilson, N. P. *et al.* Interlayer electronic coupling on demand in a 2D magnetic semiconductor. *Nature Materials* **20**, 1657-1662 (2021). https://doi.org:10.1038/s41563-021-01070-8